# The Quantum Coherence in Terms of Phase-Sensitive Nonadiabatic Dressed States

I. G. Koprinkov

*Department of Applied Physics, Technical University of Sofia, Sofia 1000, Bulgaria*

*igk@tu-sofia.bg*

**Abstract.** The quantum coherence is considered within phase-sensitive nonadiabatic dressed states. Two types of phase correlations are found: a rapidly changing phase correlation between the real and the virtual components and a stationary phase correlation between different virtual components of these states.

## INTRODUCTION

The quantum coherence is one of the most remarkable properties of the quantum phenomena, which sets the border between the quantum and classical behavior of the matter. As such, the quantum coherence becomes closely related to other fundamental quantum phenomena as quantum superposition, quantum interference, quantum entanglement, quantum decoherence, etc. [1-5]. In addition, the quantum coherence is at the heart of application of quantum physics in the quantum information science [6], quantum cryptography [7], etc. A growing interest toward a resource theory of quantum coherence has been demonstrated in relation to its technological applications [8].

The quantum coherence encompasses different concepts in the different fields of quantum physics [1]. In quantum mechanics, the simultaneous and coherent superposition of quantum states can be considered on the basis of the quantum superposition principle, which holds due to the linearity of the Schrödinger equation. The quantum coherence is a signature of definite correlations between the states of the quantum system and represents the ability of the quantum states to interfere. Such correlations are usually described by the off diagonal elements of the density matrix. The coherence of quantized bosonic/optical fields is considered as existence of definite correlations upon photon detection coincidence, *e.g.*, the Hanbury Brown and Twiss effect [9]. These are additional, intensity correlations, to the ordinary optical phase correlations. The coherence of quantized bosonic fields can be treated within the Glauber–Sudarshan coherent states [10, 11], which are eigenstates of the annihilation operator and represent the minimal uncertainty states.

In this work, the quantum coherence will be considered based on the most fundamental quantum mechanical entity, *i.e.*, the wave function. To meet the requirement of simultaneity and coherence of the quantum superposition, the problem will be treated in terms of *phase-sensitive nonadiabatic dressed states* (PSNADS) [12-14]. Two types of phase correlations are found: (*i*) a rapidly changing phase correlation between the real and virtual components and (*ii*) a stationary phase correlation between the different virtual components within given multi-level PSNADS.

## THE QUANTUM COHERENCE AND PHASE-SENSITIVE NONADIABATIC DRESSED STATES

The PSNADS allow tracing the evolution of the dynamical phase, *i.e.*, the material phase (MP), of wave function of a quantum system involved in a definite physical process, *e.g.*, interaction with an electromagnetic field and the environment.

### Physical Properties of Phase-Sensitive Nonadiabatic Dressed States

The PSNADS arise from a closed form (nonadiabatic and non-perturbative) solution of the Schrödinger equation for a two-level quantum system with a Hamiltonian (in standard notations) [12]:

$$\hat{H} = \sum_{j=1}^{2} \hbar\omega_j |j\rangle\langle j| - \mu E(t)\left(|1\rangle\langle 2| + h.c.\right) - i\hbar \sum_{j=1}^{2} \left(\gamma_j/2\right)|j\rangle\langle j| \quad , \tag{1}$$

where all phase contributions are taken into account, including the constant initial phases $\varphi_g$ and $\varphi_e$ of the ground $|1\rangle \equiv |g\rangle$ and excited $|2\rangle \equiv |e\rangle$ bare states (BS), respectively, from which the PSNADS originate due to the above mentioned interactions. The total MP of the real (index r) and virtual (index v) components of the ground (index G) and excited (index E) PSNADS, Fig.1, *e.g.*, at ground state initial conditions, are [12]:

$$\Phi_{G,r} = \varphi_g + \int_0^t \tilde{\omega}'_G \, dt' \tag{2a}$$

$$\Phi_{G,v} = \Phi_{G,r} + \Phi_F = \varphi_g + \varphi(t) + \int_0^t (\tilde{\omega}'_G + \omega) dt' \tag{2b}$$

$$\Phi_{E,r} = \Phi_{G,v} + \Phi_{NAD} = \varphi_g + \varphi(t) + \int_0^t \tilde{\omega}'_E \, dt' \tag{2c}$$

$$\Phi_{E,v} = \Phi_{E,r} - \Phi_F = \varphi_g + \int_0^t (\tilde{\omega}'_E - \omega) dt' \tag{2d}$$

where $\tilde{\omega}'_G$ and $\tilde{\omega}'_E$ are the real components Bohr frequencies/energies of ground and excited PSNADS (Stark shifted and modified by the nonadiabatic factors from the field and damping rates $\gamma_g$ and $\gamma_e$), $\Phi_F(t) = \varphi(t) + \omega t$ is the field phase, $\Delta\tilde{\omega}'_{NAD} = \tilde{\omega}'_E - \tilde{\omega}'_G - \omega$ is nonadiabatic frequency detuning, and $\Phi_{NAD} = \int_0^t \Delta\tilde{\omega}'_{NAD} dt'$ is the phase contribution due to the nonadiabatic transition between ground and excited PSNADS. Eqs.(2) show that *the MP behaves as an additive dynamical quantity which causally follows the physical processes and the initial conditions.*

Closed form solution of the PSNADS is known only for the two-level case. In order to better understand the quantum coherence, multilevel PSNADS have to be considered. The multilevel PSNADS will be obtained extrapolating the two-level case in the following way. For, *e.g.*, ground PSNADS (at ground state initial conditions), if we take the original ground BS $|g\rangle$ and repeat the procedure of derivation of PSNADS, as in [12], in a consecutive order with every one of the (electric-dipole allowed) excited BS $|e\rangle_1$, $|e\rangle_2$, $|e\rangle_3$ ..., the result will be generally same, as in the two-level PSNADS, with the difference that for each excited BS (due to the different detuning $\Delta\omega_i$ from the exact resonances, Fig.2, and the dipole moment matrix elements between the states), the strengths of the respective virtual states, denoted by $C_{vi}$, $i = 1, 2, 3..$, (instead of $SIN(\theta/2)$ for the two-level case [12]), and the dynamic Stark shift will be different. If we now allow all these excited BS to act simultaneously, as in the reality, each excited BS $|e\rangle_1$, $|e\rangle_2$, $|e\rangle_3$ ..., while evolving to respective real component $|E_r\rangle_1$, $|E_r\rangle_2$, $|E_r\rangle_3$ ..., will contribute to the creation of a respective virtual component $|G_v\rangle_1$, $|G_v\rangle_2$, $|G_v\rangle_3$ ... of the ground PSNADS, Fig.2. At the same time, the ground BS $|g\rangle$ of Bohr frequency/energy $\omega_g$ evolves toward a single only (non-degenerate case) real component $|G_r\rangle$ of the ground PSNADS of a well-defined Bohr frequency/energy $\tilde{\omega}'_{G,r} \equiv \tilde{\omega}'_G$. The carrier frequency (photon energy) $\omega$ of the field also keeps a well-defined value (at absent, as usual in these studies, of nonlinear processes). As the Bohr frequency/energy of each virtual component is equal to the Bohr frequency/energy of the (single) real ground state $\tilde{\omega}'_{G,r} \equiv \tilde{\omega}'_G$ plus one photon energy $\omega$, all virtual components $|G_v\rangle_1$, $|G_v\rangle_2$, $|G_v\rangle_3$ ..., will have same Bohr frequency/energy: $\tilde{\omega}'_{G,vi} = \tilde{\omega}'_{G,r} + \omega$, $i = 1, 2, 3...$, Fig.2., and, thus, same rate of MP acquisition in time. If the factors that determine the time dependent phase acquisition of the different virtual components $|G_v\rangle_i$ are equal, the only phase difference between different virtual components may arise, eventually, from the appearance of a constant phase shift $\varphi_{G,v,i}$, $i = 1, 2, 3...$, for each virtual component $|G_v\rangle_i$ in the process of its creation.

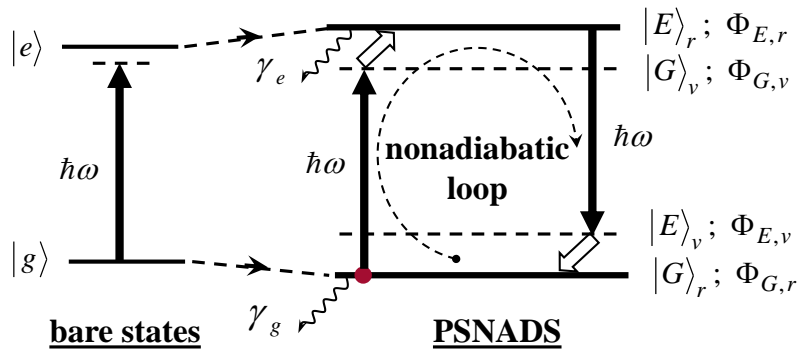

Fig.1. Material phase of the real and virtual components of ground and excited PSNADS (two-level case)

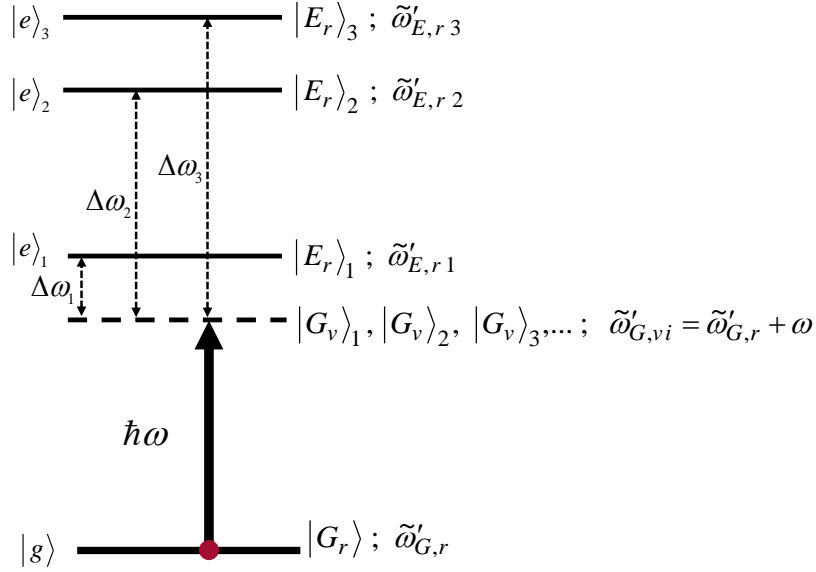

Fig.2. Multi-level PSNADS. The initial BS (left), the real components (full lines), and the virtual components (broken line) of the ground PSNADS together with the Bohr frequencies/energies of the states are shown at the right of the figure.

Based on the above considerations, the following conclusions about the multi-level PSNADS can be done. Each PSNADS consists of a single real component and a number of virtual components, equal to the number of the initial electric-dipole allowed BS involved in the creation of PSNADS. Thus, the structure of, e.g., ground PSNADS, is:

$$|G\rangle = C_r |G_r\rangle + \sum_i C_{vi} |G_v\rangle_i \quad , \quad i = 1, 2, 3... \quad . \quad (3)$$

The strengths (superposition coefficients) of the real and virtual components, $C_r$ and $C_{vi}$ respectively, of the multi-level case stand for the respective strengths, $COS(\theta/2)$ and $SIN(\theta/2)$, of the two-level case [12]. Similar structure will also have multi-level excited PSNADS, but, beside of a single real component, it will consist of upward and downward virtual components due to upward and downward excitations by the field. The strength, $SIN(\theta/2)$, of the virtual component is zero at zero-field and nonzero at nonzero-field, as in the case of nonadiabatic dressed states NADS [15]. Extrapolating these properties toward the multi-level case, one may conclude that the virtual components appear simultaneously from the respective real component of given PSNADS with switching the field on and disappear with switching the field off. The virtual components cannot exist independently of the respective real component, from which they originate, and the forcing electromagnetic field. Therefore, we consider (within the concept of quantum state) that the superposition of real and virtual components of given PSNADS is *simultaneous* within the overlap time of the real component lifetime and the time of action of the field that creates the virtual components. This is in strong contrast to the superposition of the most widely used BS, which, as has been shown earlier [16], cannot be superimposed simultaneously and coherently.

The total phases of the real and the virtual components of the multi-level ground PSNADS, after modification of Eqs. (2a) and (2b), i.e., adding a constant phase $\varphi_{G,v,i}$ to each virtual component, as it is specified above, are:

$$\Phi_{G,r} = \varphi_g + \int_0^t \tilde{\omega}'_G \, dt' \qquad (4a)$$

$$\Phi_{G,v,i} = \varphi_{G,v,i} + \varphi_g + \varphi(t) + \int_0^t (\tilde{\omega}'_G + \omega) \, dt' \quad , \quad i = 1, 2, 3... \quad . \quad (4b)$$

In agreement with the adiabatic theorem of quantum mechanics, transitions between different states result from nonadiabatic factors acting on the quantum system. The nonadiabatic factors have, in general, a stochastic nature due to, at least, the zero-point vacuum fluctuations, which act universally on any quantum system and cannot be removed. This destroys the quantum coherence between different PSNADS. Consequently, strictly speaking, *quantum coherence may exist only within given PSNADS*. Transitions between different PSNADS, *e.g.*, from ground to excited PSNADS, lead to stochastic phase contributions to the excited PSNADS. Similar coherence behavior will exist if the process develops at excited state initial conditions. Within the present considerations, we will focus our attention on the coherence of the ground PSNADS.

The field creating the PSNADS is, in general, non-monochromatic and the virtual components of the PSNADS will have a nonzero spectral/energy width. In this case, the superposition of virtual components may lead to formation of localized material wave-packets inside the quantum system (atom, molecule, etc.), which allows tracing its internal dynamics.

## The Quantum Coherence Explained Within Phase-Sensitive Nonadiabatic Dressed States

The spectral properties of the PSNADS, *i.e.*, same Bohr frequency/energy and a constant phase shift of the different virtual components, have a fundamental importance for the quantum coherence in terms of quantum states and for the interference of quantum states, as well. Another important point for the quantum interference is the simultaneity in the superposition of the quantum states.

Seeking for quantum coherence of the states, the phase differences between (*i*) the virtual and the real components, Eq.(5a), and (*ii*) between the different virtual components of given PSNADS (in this case - the ground PSNADS), Eq.(5b), will be considered based on Eqs.(4):

$$\Phi_{G,v,i} - \Phi_{G,r} = \varphi_{G,v,i} + \varphi(t) + \omega t \qquad i = 1, 2, 3... \qquad (5a)$$

$$\Phi_{G,v,i} - \Phi_{G,v,j} = \varphi_{G,v,i} - \varphi_{G,v,j} = const \qquad i, j = 1, 2, 3... \qquad (5b)$$

Two types of correlations can be distinguished from Eqs.(5):
(*i*): "*fast*"/"*hidden*" correlation or "*low coherence*", Eq.(5a).
(*ii*) "*slow*" correlation or "*high coherence*", Eq.(5b).
As can be seen from Eq.(5a), a well-defined regular phase difference exists between the phases of the virtual and the real components of the PSNADS, nevertheless it rapidly changes in time due to the large phase acquisition term $\omega t$, where $\omega$, beside of the carrier frequency of the field, represents the frequency/energy difference between the virtual and the real components, $|G_v\rangle_i$ and $|G_r\rangle$, respectively, Fig.2. Interference of this type is hard to be observed experimentally because the superposition distribution of $|G_v\rangle_i$ and $|G_r\rangle$ components change in time with a huge rate in the order of the optical frequency $\omega$. Therefore, it will be considered as a "hidden" coherence. On the other hand, Eq.(5b) shows an existence of a constant phase difference between any two virtual components $|G_v\rangle_i$ without rapidly oscillating term because all virtual components have same Bohr frequency/energy $\widetilde{\omega}'_{G,vi} = \widetilde{\omega}'_{G,r} + \omega$. This type of coherence leads to a stable and a well (but not easy) observable quantum interference distribution between the different virtual components $|G_v\rangle_i$ of the PSNADS. Therefore, it will be considered as a "high coherence".

The following *definition of quantum coherence* will be introduced based on the properties of "*high coherence*": *two or more (simultaneously superimposed) quantum states are coherent if they have same energy (Bohr frequency) and a constant difference between phases of these states*. The present approach allows generalizing the notion of "coherence" putting on the same footing the quantum coherence and the ordinary coherence of optical fields.

## Quantum Coherence and Dynamics-Statistical Interpretation of Quantum Mechanics

The present approach goes out of the frames of the most widely accepted interpretation of quantum mechanics - the Copenhagen interpretation (CI). The CI confers physical meaning to the amplitude of the wave function only but not to the phase. It, however, can be well accommodated in the Dynamics-Statistical Interpretation (DSI) of quantum mechanics [14]. According to the DSI both elements of the wave function, the amplitude and the phase, are causally related to the physical reality due to the following types of arguments [14].

(*i*) *Special theoretical argument* – based on the behavior of the MP within the PSNADS [12]. The MP of the PSNADS (as any other physical quantity) depends causally on the physical processes and the initial conditions.

(*ii*) *General theoretical argument* – based on the hydrodynamic representation of quantum mechanics [17]. The hydrodynamic representation of quantum mechanics shows that the amplitude and the phase of the wave function are not independent but codetermine each other because they obey coupled differential equations. Consequently, the well-known relation of the amplitude of wave function to the physical reality, as in the CI, leads to a definite relation of the phase of the wave function to the very same physical reality.

(*iii*) *Experimental evidences* – based on experiments with material wave packets within atoms [18] and molecules [19]. The dependence of observable quantities, *e.g.*, the population of a given quantum state, on the MP in the interference of intra-atomic and intra-molecular wave packets is proved experimentally. Not only the time dependent but even the constant change of the MP leads to observable results in these experiments.

## CONCLUSIONS

The phase coherence between virtual and real components as well as between different virtual components, superposed simultaneously within given PSNADS, is found. The PSNADS are a natural frame to treat the quantum coherence directly in terms of quantum states. The quantum coherence in terms of quantum states can be well understood within the dynamics-statistical interpretation of quantum mechanics.

**Acknowledgements: The author would like to thank to the Research and Development Sector at the Technical University of Sofia for the financial support.**